\documentclass[12pt,a4paper]{article}
\usepackage{graphics}
\newcommand\smfrac[2]{{\textstyle{\frac{#1}{#2}}}}

\begin{document}
\begin{titlepage}
\begin{flushright}
  BICOCCA-FT-99-38\\
  RAL-TR-1999-076 \\
  hep-ph/9911335
\end{flushright}
\par \vspace{10mm}
\begin{center}
{\Large \bf
Simulations of Top Production and Decay \\[1ex]
at the Linear Collider\footnote{Talk given by G. Corcella at the "2nd 
ECFA/DESY Study on Physics and Detectors for a Linear Electron--Positron
Collider", LNF, Frascati, Italy, 8-10 November 1998.}}
\end{center}
\par \vspace{2mm}
\begin{center}
{\bf G. Corcella $^1$\footnote{Address after the 1st of November 1999:
Department of Physics and Astronomy, University of Rochester, Rochester, NY 
14627, U.\ S.\ A.}
 and M.H. Seymour $^2$}\\
\vspace{2mm}
{$^1$ Dipartimento di Fisica, Universit\`a di Milano
  and INFN, Sezione di Milano}\\
{Via Celoria 16, I-20133, Milano, Italy}
\par \vspace{2mm}
\vspace{2mm}
{$^2$ Rutherford Appleton Laboratory, Chilton,}\\
{Didcot, Oxfordshire.  OX11 0QX\@.  U.K.}
\end{center}
\par \vspace{2mm}
\begin{center} 
{\large \bf Abstract} 
\begin{quote}
\pretolerance 10000
{We review the present status of simulations of top production and decay
in the HERWIG Monte Carlo event generator. 
We show the phenomenological impact of the recently-implemented 
matrix-element corrections to top decays 
for $e^+e^-$ collisions at 360 GeV and discuss possible further improvements
for studies of future experiments at the Linear Collider.}
\end{quote}
\end{center}
\vspace*{\fill}
\begin{flushleft}
  RAL-TR-1999-076\\
  BICOCCA-FT-99-38\\
  November 1999 
\end{flushleft}
\end{titlepage}

\section{Introduction}
The study of the top quark phenomenology will be among the main topics of 
interest for the future experiments at the Linear Collider. In order to 
perform such analyses, it will be 
essential to have reliable Monte Carlo simulations of
top production and decay.
Standard event generators [\ref{pythia},\ref{herwig}] simulate parton 
showers in the soft/collinear approximation, leaving empty regions in 
the phase space (`dead zones') where, according to the exact matrix element, 
the radiation should be suppressed but not completely absent.
Parton showers need therefore to be supplemented by matrix-element corrections.

As far as the HERWIG parton shower is concerned, matrix-element corrections to 
$e^+e^-$ annihilation [\ref{sey1}], deep inelastic scattering 
[\ref{sey2}], top decays [\ref{corsey1}] and Drell--Yan processes
[\ref{corsey2}] have been implemented 
following the general prescriptions
given in [\ref{sey3}].
In section 2 we shall review the HERWIG parton shower algorithm for
top production and decay in $e^+e^-$ collisions and discuss the recent 
implementation of matrix-element corrections to top decays.
In section 3 we shall show some phenomenological results at the Linear 
Collider, for a centre-of-mass energy of 360 GeV.
In section 4 we shall make some concluding remarks on the results
and possible improvements of this work.

\section{The HERWIG parton shower algorithm for top production and decay
in $e^+e^-$ collisions}
For the emission of one more parton (gluon) in top production in
$e^+e^-$ annihilation
($e^+e^-\to t\bar t g$) and top decay ($t\to bWg$), the elementary probability 
implemented in the HERWIG event generator is the general result for parton 
radiation in the soft/collinear limit [\ref{marweb1}]:
\begin{equation}
\label{elementary}
  dP={{dq^2}\over{q^2}}\;{{\alpha_S}\over {2\pi}}\;P(z)dz\;
  {{\Delta_S(q^2_{\mathrm{max}},q^2_c)}\over{\Delta_S(q^2,q^2_c)}}\ .
\end{equation}
The shower is ordered according to the variable $q^2=E^2\xi$, $E$ being the 
energy of the parton that splits and $\xi=p_1\cdot p_2/(E_1E_2)$, where $p_1$ 
and $p_2$ are the momenta of the two outgoing partons; $z$ is the energy 
fraction of the emitted parton (gluon) relative to the incoming one.
In the massless approximation, $\xi=1-\cos\theta$, $\theta$ being the 
emission angle, 
in such a way that ordering according to $q^2$, in the soft limit, results in 
angular ordering.
$\Delta_S(q^2,q^2_c)$ is the Sudakov form factor, expressing the probability 
that no resolvable radiation is emitted from a parton whose upper limit on
emission is $q^2$, with $q_c^2$ being a cutoff on transverse momentum. 
The ratio of Sudakov form 
factors in (\ref{elementary}) sums up all virtual contributions and unresolved 
emissions.

The parton shower variables $z$ and $\xi$ are frame-dependent, 
nevertheless it is possible to prove that colour coherence implies that the
values $q_{\mathrm{max}}$ of any pair of colour-connected partons are related 
via $q_{i\mathrm{max}}q_{j\mathrm{max}}=p_i\cdot p_j$, which is 
Lorentz-invariant.
For top production, as for most of the HERWIG processes, symmetric limits are 
chosen, i.e. $q_{i\mathrm{max}}=q_{j\mathrm{max}}=\sqrt{p_i\cdot p_j}$
and the energy of the parton which initiates the cascade is equal to
$E_i=q_{i\mathrm{max}}$. Ordering according to $q^2$ leads to the
condition $\xi<1$.
The phase space region $1<\xi<2$ is not populated at all by the 
parton shower algorithm. In [\ref{sey1}] matrix-element corrections are 
implemented to $e^+e^-\to q\bar q$ processes: 
the `dead zone' of the phase space is 
populated according to a probability distribution obtained
from the exact ${\cal O}(\alpha_S)$ matrix-element calculation
of the process 
$e^+e^-\to q\bar q g$ (`hard correction') and the emission in the 
already-populated region is corrected using the exact amplitude for 
every emission that is the hardest so far (`soft correction').

Top decays are treated in a somewhat different way in HERWIG, since the top 
quark rest frame is chosen to perform such a decay [\ref{marweb2}]. 
This means that 
$E_t=m_t=q_{t\mathrm{max}}$ and $E_b=p_t\cdot p_b/m_t$. Being at rest,
the top quark is not allowed to radiate gluons in the decay stage,
while the $b$ quark emits soft gluons for
$\xi<1$, i.e. $0<\theta<\pi/2$.
The result is that the soft phase space is not completely populated by HERWIG 
as all the soft gluons which should be emitted in the backward hemisphere 
$\pi/2<\theta <\pi$
are missed. For $e^+e^-$ annihilation we have two back-to-back partons,
$q$ and $\bar q$, that are capable of emitting soft gluons for 
$\xi<1$, in such a 
way that each of them radiates in the region the other is missing out
and the whole of the soft phase space is filled.
In [\ref{marweb2}] it is shown that, though neglecting the soft radiation
in the backward hemisphere, the total energy loss due
to gluon radiation is roughly right. However, we may have serious problems when 
dealing with angular differential distributions.

For the reasons previously mentioned, the implementation of matrix-element 
corrections to top decays is 
not a straightforward extension of the method applied for the
purpose of other processes. In order to avoid the soft singularity, we set a 
cutoff on the energy of backward gluons, whose default value is 
$E_{\mathrm{min}}=2$~GeV.
In [\ref{corsey1}] we calculated the phase space limits for 
the decay $t(q)\to b(p_2) W(p_1) g(p_3)$ with respect to the variables
\begin{eqnarray}
x_1&=&1-{{2p_2\cdot p_3}\over{m_t^2}},\\
x_3&=&{{2p_3\cdot q}\over {m_t^2}}.
\end{eqnarray}
\begin{figure}[t]
\centerline{\resizebox{0.50\textwidth}{!}{\includegraphics{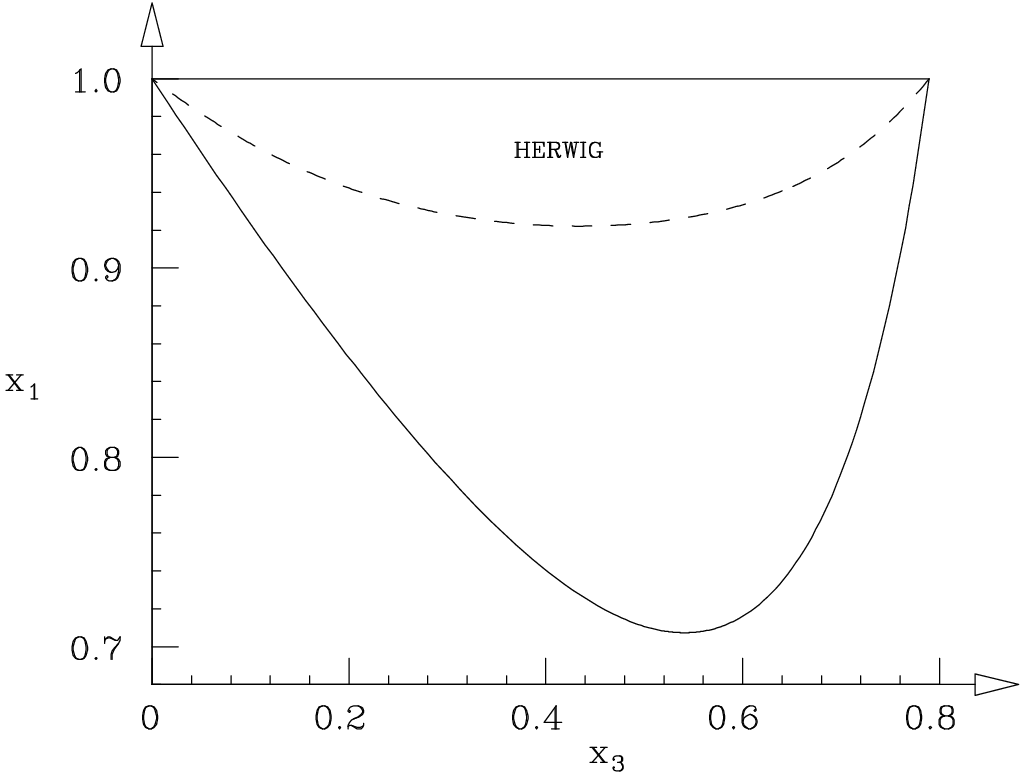}}}
\caption{The total and HERWIG phase space in terms of $x_1$ and $x_3$.
The soft singularity $x_3=0$ is not completely inside the HERWIG region.}
\label{phase}
\end{figure}
In Fig.~\ref{phase} we plot the total and the HERWIG phase space.
The differential width reads:
\begin{eqnarray}\label{gamma}
  \frac1{\Gamma_0}\frac{d^2\Gamma}{dx_1\,dx_3} &=&
  \frac1{(1-a)
  \left( 1 + \frac1a - 2a \right)}
  2 \;
  \frac{\alpha_s}{2\pi}
  \, C_F
  \frac1{x_3^2(1-x_1)}\Biggl\{\scriptstyle
 (1+\smfrac1a-2a)
 \left[(1-a)x_3-(1-x_1)(1-x_3)-x_3^2\right]
\nonumber\\&&\scriptstyle
 +(1+\smfrac1{2a})x_3(x_1+x_3-1)^2
 +2x_3^2(1-x_1)
 \Biggr\},
\end{eqnarray}
where $\Gamma_0$ is the width of the Born process $t\to bW$ and
$a=m_W^2/m_t^2$.
The integral of the differential width over the `dead zone'
is divergent, because of the soft divergence.
We use the differential width (\ref{gamma}) to generate events in the missing 
backward phase space and in the already-populated forward hemisphere for every hardest-so-far 
emission capable of being the hardest so far. 
Both hard and soft corrections should be applied only for gluon 
energies larger than the cutoff value.

\section{Phenomenological results at the Linear Collider}
In order to test the reliability of the HERWIG algorithm provided 
with matrix-element corrections we wish to compare its phenomenological 
results with the ones obtained by calculating the exact first-order 
matrix element of the process: 
\begin{equation}
e^+e^-\to t\bar t\to (bW^+)(bW^-)g.
\end{equation}
It is interesting to perform such an analysis at the centre-of-mass energy 
$\sqrt{s}=360$~GeV, slightly above the threshold for $t\bar t$
production, in such
a way that all the gluon emission is associated to top decays, 
the available phase space for radiation in the production stage being 
too small. 
In the last public version HERWIG 5.9 some bugs were found in
the treatment of 
top decays, therefore, if we wish to investigate the effect of the implemented 
matrix-element corrections, it is better to compare the new 
version HERWIG 6.1 to an intermediate version, which we call HERWIG 6.0, where 
all the bugs are supposed to be fixed.
\begin{figure}[t]
\centerline{\resizebox{0.50\textwidth}{!}{\includegraphics{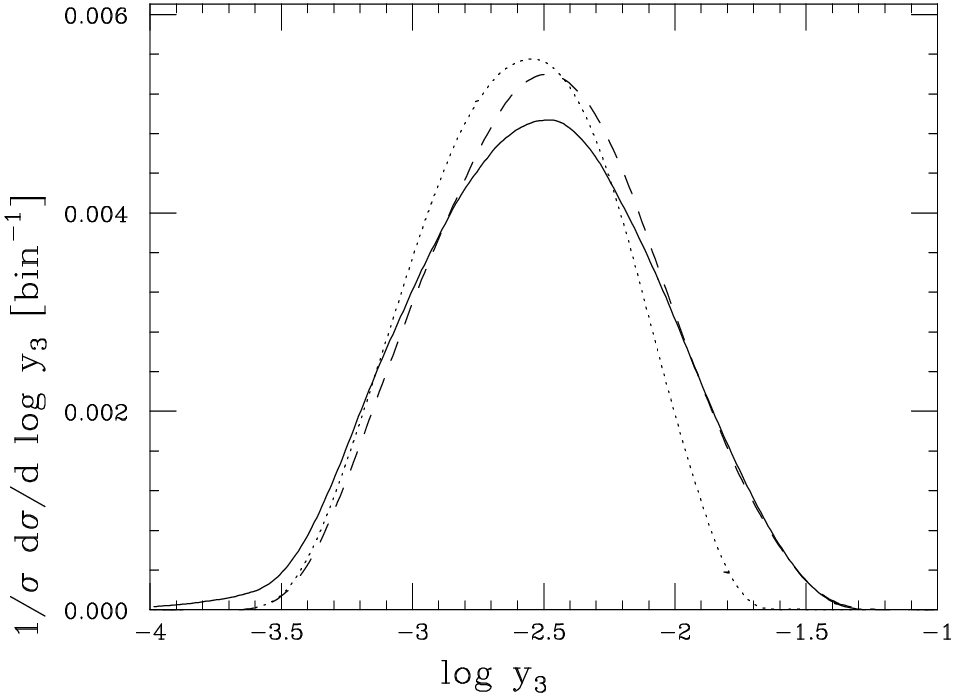}}}
\caption{$y_3$ distributions according to HERWIG 6.1 (dashed line),
HERWIG 6.0 (dotted) and according to the exact ${\cal O}(\alpha_S)$ calculation
(solid).}
\label{y3}
\end{figure}
\begin{figure}
\centerline{\resizebox{0.50\textwidth}{!}{\includegraphics{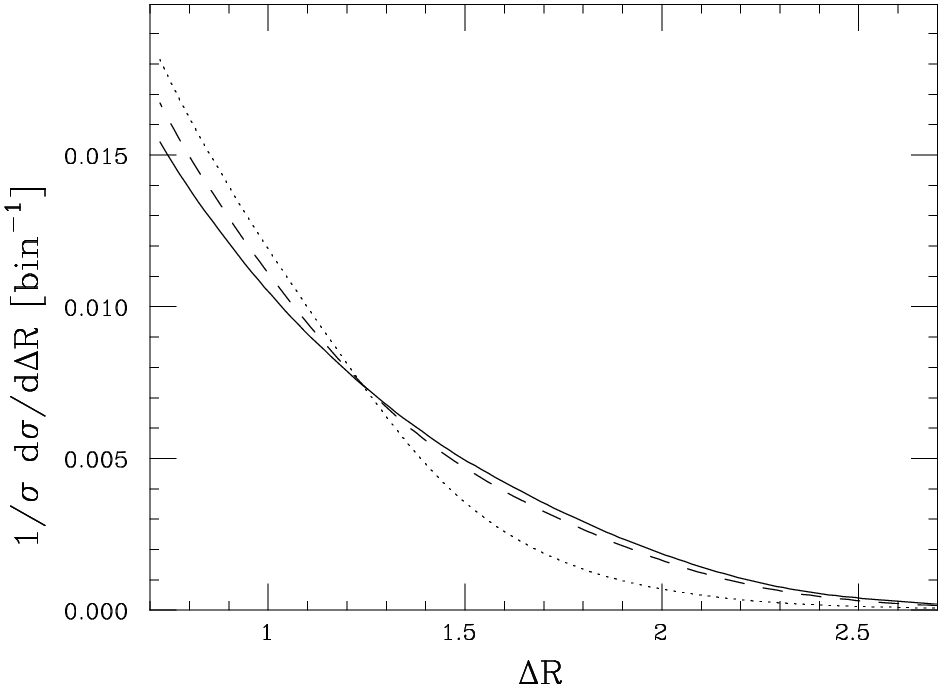}}}
\caption{Invariant opening angle distributions
according to HERWIG 6.1 (dashed line), HERWIG 6.0 (dotted line)
and according to the exact ${\cal O}(\alpha_S)$ matrix element (solid).}
\label{deltar}
\end{figure}

We study three-jet events according to the Durham $k_T$ algorithm [\ref{kt}]
and plot the differential distributions with respect to the threshold 
variable $y_3$ and the minimum invariant opening angle 
$\Delta R=\sqrt{(\Delta\phi)^2+(\Delta\eta)^2}$ using HERWIG 6.1 and 6.0
and the exact ${\cal O}(\alpha_S)$ for a value $\alpha_S=0.145$ of the strong
coupling constant. We also set the cuts $E_T>10$~GeV and $\Delta R>$~0.7 on
the resolved jets. Such an analysis was already performed in [\ref{orr}] and 
serious discrepancies were found when comparing the exact results
with the ones obtained by the using of HERWIG before the implementation of 
matrix-element corrections to top decays.

In Figs.~\ref{y3} and \ref{deltar} we observe
a remarkable impact of matrix-element corrections:
HERWIG 6.1 generates more events at large $y_3$ with respect to the 6.0 
version, which can be explained as due to the
hard corrections which allow hard and large-angle gluon 
radiation. We also get a suppression at small $y_3$, which compensates the
enhancement at large $y_3$, the total amount of radiation being roughly
right even before matrix-element corrections. The soft corrections, applying
the matrix-element distribution instead of the parton shower one in the 
already-populated forward region for every hardest-so-far emission, 
is responsible of the suppression at small $y_3$.
The agreement between HERWIG 6.1 and the exact 
calculation is pretty good at large $y_3$, as it should be since 
the fixed-order calculation is reliable only if we are far from the
soft and collinear divergences, which correspond to small $y_3$, 
where, on the contrary, parton showers are more trustworthy.
Similar comments hold for the $\Delta R$ plots as well.
In [\ref{corsey1}] we also showed that the dependence on the chosen value of
the cutoff
$E_{\mathrm{min}}$ is negligible after we apply the experimental cuts.

\section{Conclusions}
We have reviewed the HERWIG parton shower model for top production and decay 
in $e^+e^-$ annihilation 
and the main features of the method of matrix-element corrections, recently 
implemented to the simulation of top decays.
We studied three-jet events at $\sqrt{s}=360$~GeV 
using HERWIG before and after matrix-element corrections to top decays,
and the exact first-order calculation.
The results showed a marked effect of the improvement to the treatment of top
decays and a good agreement with the ${\cal O}(\alpha_S)$ calculation
in the phase space region where such an agreement is to be expected.
We therefore feel confident that the new version 
HERWIG 6.1 should be a reliable Monte Carlo event generator to 
simulate top quark decay at the future Linear Collider. 

For the sake of completeness, we have however 
to say that the implementation of matrix-element 
corrections to the process $e^+e^-\to q\bar q g$ still needs some improvement 
since, in the determination of the phase space limits and in the application of 
the soft corrections, the $q\bar q$ pair is treated as if it was massless, 
which might not be a good approximation for the purpose of the top quark.
We have been working on fully including these mass effects.
Furthermore, it will be very interesting to investigate the impact of 
matrix-element corrections to top decays on the top mass reconstruction
at the Linear Collider. This is in progress as well.

\section*{References}
\begin{enumerate}
\addtolength{\itemsep}{-0.5ex}
\item\label{pythia}
 T. Sj\"ostrand, Comp.\ Phys.\ Comm.\ 46 (1987) 367.
\item\label{herwig}
  G. Marchesini et al.\ Comput.\ Phys.\ Commun.\ 67 (1992) 465
\item\label{sey1}
 M.H. Seymour, Z.\ Phys.\ C56 (1992) 161.
\item\label{sey2}
 M.H. Seymour, {\it Matrix Element Corrections to Parton Shower
    Simulation of Deep Inelastic Scattering}, contributed to 27th
  International Conference on High Energy Physics (ICHEP), Glasgow,
  1994, Lund preprint LU-TP-94-12, unpublished.
\item\label{corsey1}
 G. Corcella and M.H. Seymour, Phys.\ Lett.\ B442 (1998) 417.
\item\label{corsey2}
 G. Corcella and M.H. Seymour, hep-ph/9908388, accepted for publication in 
Nucl. Phys. B.
\item\label{sey3}
 M.H. Seymour, Comput.\ Phys.\ Commun.\ 90 (1995) 95.
\item\label{marweb1}  
 G. Marchesini and B.R. Webber, Nucl.\ Phys.\ B238 (1984) 1.
\item\label{marweb2}
 G. Marchesini and B.R. Webber, Nucl.\ Phys.\ B330 (1990) 261.
\item\label{kt}
 S. Catani, Yu.L. Dokshitzer, M. Olsson, G. Turnock and B.R. Webber, Nucl.\ 
Phys.\ B406 (1991) 432.
\item\label{orr}
 L.H. Orr, T. Stelzer and W.J. Stirling, Phys. Rev. D 56 (1997) 446.

\end{enumerate}
\end{document}